\documentclass[conference,letterpaper]{IEEEtran}

\usepackage[utf8]{inputenc}
\usepackage[T1]{fontenc}
\usepackage{url}
\usepackage{ifthen}
\usepackage{cite}
\usepackage[cmex10]{amsmath}
\usepackage{amssymb}
\usepackage{graphicx}
\usepackage[subfigure]{graphfig}
\usepackage{amsmath}

\begin{document}
\title{On Massive IoT Connectivity with Temporally-Correlated User Activity}


\author{%
  \IEEEauthorblockN{Qipeng Wang, Liang Liu, Shuowen Zhang, and Francis C. M. Lau}
  \IEEEauthorblockA{Department of Electronic and Information Engineering\\
                    The Hong Kong Polytechnic University\\
                    Emails:
                    qipeng.wang@connect.polyu.hk} $\{$liang-eie.liu,shuowen.zhang,francis-cm.lau$\}$@polyu.edu.hk}


\maketitle

\begin{abstract}
   This paper considers joint device activity detection and channel estimation in Internet of Things (IoT) networks, where a large number of IoT devices exist but merely a random subset of them become active for short-packet transmission at each time slot. In particular, to improve the detection performance, we propose to leverage the \emph{temporal correlation} in user activity, i.e., a device active at the previous time slot is more likely to be still active at the current time slot. Despite the appealing temporal correlation feature, it is challenging to unveil the connection between the estimated activity pattern for the previous time slot (which may be imperfect) and the true activity pattern at the current time slot due to the unknown estimation error. In this paper, we manage to tackle this challenge under the framework of approximate message passing (AMP). Specifically, thanks to the state evolution, the correlation between the activity pattern estimated by AMP at the previous time slot and the real activity pattern at the previous and current time slot is quantified explicitly. Based on the well-defined temporal correlation, we further manage to embed this useful SI into the design of the minimum mean-squared error (MMSE) denoisers and log-likelihood ratio (LLR) test based activity detectors under the AMP framework. Theoretical comparison between the SI-aided AMP algorithm and its counterpart without utilizing temporal correlation is provided. Moreover, numerical results are given which show the significant gain in activity detection accuracy brought by the SI-aided algorithm.
\end{abstract}


\section{Introduction}\label{sec:Introduction}

A typical massive Internet of Things (IoT) connectivity system consists of a large number of low-cost devices, each of which stays in the silence mode for a long period to save the energy and becomes active merely when triggered by the unusual events. Under such a setting, one key challenge lies in how to jointly identify the randomly active IoT devices and estimate their channels in a fast and accurate manner \cite{mMTC1}. Recently, it was shown that the above job can be accomplished by utilizing the compressed sensing technique thanks to the sparse user activity \cite{mMTC2,mMTC3,mMTC4,mMTC5,mMTC6,mMTC7}. In particular, under the framework of multiple measurement vector (MMV) based approximate message passing (AMP) \cite{AMP,MMV_AMP}, it has been shown in \cite{mMTC2} that the activity detection error probability decreases significantly with the number of antennas at the base station (BS). Such an exciting result arises from reaping the spatial correlation in user activity: if one device is active for one antenna, it is also active for all the other antennas. However, this theoretical performance gain is achieved at the cost of high computational complexity in practice: in an IoT system with a large number of devices and BS antennas, the dimension of the data to be processed is tremendous. A nature question is: if only a small number of antennas are utilized to reduce the computational complexity, is it still possible to achieve high-quality device activity detection and channel estimation?

This paper provides an affirm answer to the above question. The core is to utilize the temporal correlation in user activity to compensate for the spatial correlation gain of the MMV-based AMP algorithm. In practice, temporally-correlated user activity may come from the fact that if an abnormal event is detected by some sensor at a moment, then this device is more likely to be still activated by this event in the near future. To fully take advantage of this temporal correlation, this paper aims to design a side information (SI) aided AMP framework.

Note that at each time slot, the available information at the BS is the imperfect device activity pattern estimated at the previous time slot, whose connection to the real device activity pattern at the previous or current time slot is unclear in general due to the unknown estimation error. As a result, despite the existence of temporal correlation in user activity, it is a challenging task to utilize this as SI to improve the performance of activity detection. In this work, we point out that under the framework of AMP, the correlation between the device activity pattern estimated at the previous time slot and the real one at the current time slot can be explicitly quantified thanks to the state evolution. Based on this correlation, we further manage to design the minimum mean-squared error (MMSE) denoisers and log-likelihood ratio (LLR) based detectors in AMP with SI taken into consideration. The impact of using SI in AMP is illustrated both theoretically and numerically.

In the literature, temporal correlation has been also utilized in \cite{DCS-AMP,SI-AMP} to design the AMP algorithm. However, different from \cite{DCS-AMP} that considers a Turbo extension of the AMP algorithm \cite{Turbo-AMP} based on the idea of factor graph, our approach provides a framework to incorporate SI into the MMV-AMP algorithm without needing to craft the graph model for each new signal. Note that \cite{DCS-AMP} merely works for the single measurement vector (SMV) case, i.e., the BS has a single antenna. More importantly, even for the special case of the SMV problem, our approach can be shown to be Bayes-optimal, thus yielding improved detection performance. Moreover, compared with \cite{SI-AMP} whose emphasize is on estimating the sparse channels, our paper places more focus on device activity detection. As a result, dedicated activity detectors are proposed based on the SI and the LLR test.

\section{System Model}\label{sec_sys}

\emph{Baseband Model:} This paper considers the uplink communication in a massive IoT connectivity system consisting of one BS equipped with $M$ antennas and $N$ single-antenna IoT devices. We assume quasi-static block-fading channels, in which all user channels remain approximately constant in each coherence block, but vary independently from block to block. Let $J$ denote the number of consecutive coherence blocks considered in this work. At coherence block $j$, the channel from device $n$ to the BS is denoted by $\boldsymbol{h}_n^{(j)}\in \mathbb{C}^{M\times 1}$, $j=1,\ldots,J$, $n=1,\ldots,N$. It is assumed that the user channels follow the independent and identically distributed (i.i.d.) Rayleigh fading channel model, i.e., $\boldsymbol{h}_n^{(j)}\in \mathcal{CN}(\boldsymbol{0},\gamma_n\boldsymbol{I})$, $\forall j,n$, where $\gamma_n$ is the path loss of device $n$. Note that $\boldsymbol{h}_n^{(j)}$'s are independent over $n$ and $j$.

Due to the sporadic data traffic in IoT networks, only a small set of devices become active in each coherence block. We define the user activity indicator functions as follows:
 \begin{equation}\label{indicator}
     \delta_n^{\left(j\right)}=
\begin{cases}
1,& \text{if user $n$ is active at coherence block} ~ j,\\
0,& \text{otherwise},
\end{cases} ~ \forall j,n,
\end{equation}so that $\delta_n^{\left(j\right)}$ is a Bernoulli random variable with
\begin{equation}\label{p_active}
Pr(\delta_n^{\left(j\right)}=1)=\lambda, ~ Pr(\delta_n^{\left(j\right)}=0)=1-\lambda, ~~~ \forall n, j.
\end{equation}

In this work, we consider the grant-free random access scheme \cite{mMTC1} in our interested IoT system, where at the beginning of each coherence block, the active devices transmit their pilot sequences to the BS to perform joint device activity detection and channel estimation. Let $\boldsymbol{s}_n=[{s}_{n,1}, \ldots, {s}_{n,L}]^T\in \mathbb{C}^{L\times 1}$ denote the pilot sequence with length $L$ assigned to device $n$, $\forall n$. Similar to \cite{mMTC1,mMTC2,mMTC3}, it is assumed that all the entries in $\boldsymbol{s}_n$ are generated from i.i.d. complex Gaussian distribution with zero mean and variance $1/L$, $\forall n$. Then, the BS received signal at coherence block $j$ is expressed as
\begin{equation}\label{basebandmodel}
\boldsymbol{Y}^{\left(j\right)}= \sum_{n=1}^{N} \delta_n^{\left(j\right)}\boldsymbol{h}_n^{\left(j\right)}\boldsymbol{s}_n+\boldsymbol{Z}^{\left(j\right)}=\boldsymbol{S}\boldsymbol{X}^{\left(j\right)}+\boldsymbol{Z}^{\left(j\right)}, ~~~ \forall j,
\end{equation}
where $\boldsymbol{Z}^{\left(j\right)} \in \mathbb{C}^{L\times M} \in \mathcal{CN}(\boldsymbol{0},\sigma_z^2\boldsymbol{I})$ is the additive white Gaussian noise (AWGN) of the BS at coherence block $j$, $\boldsymbol{S}=[\boldsymbol{s}_1,\ldots,\boldsymbol{s}_N]\in\mathbb{C}^{L\times N}$, and $\boldsymbol{X}^{\left(j\right)}=[\boldsymbol{x}_1^{\left(j\right)}, \ldots,\boldsymbol{x}_N^{\left(j\right)}]^T \in\mathbb{C}^{N\times M}$ with $\boldsymbol{x}_n^{\left(j\right)}=\delta_n^{\left(j\right)}\boldsymbol{h}_n^{\left(j\right)}$ denoting the effective channel of device $n$ at coherence block $j$, $\forall j,n$. At each coherence block $j$, the job of the BS is to jointly detect the active devices and estimate their channels by estimating $\boldsymbol{X}^{\left(j\right)}$ based on its received signal $\boldsymbol{Y}^{\left(j\right)}$ and its knowledge of the user pilots $\boldsymbol{S}$.

\emph{Temporally-Correlated User Activity Model:} This paper considers the case of temporally-correlated user activity, which is modeled by a Markov chain with the following transition probabilities:
\begin{equation}\label{p_transition}
\begin{aligned}
Pr(\delta_n^{\left(j\right)} = 1\mid \delta_n^{\left(j-1\right)}=1)&=\alpha,\\
Pr(\delta_n^{\left(j\right)} = 0\mid \delta_n^{\left(j-1\right)}=1)&=1-\alpha,\\
Pr(\delta_n^{\left(j\right)}= 1\mid \delta_n^{\left(j-1\right)}=0)&=\beta,\\
Pr(\delta_n^{\left(j\right)} = 0\mid \delta_n^{\left(j-1\right)}=0)&=1-\beta,
\end{aligned}~~~ \forall j,n.
\end{equation}In other words, if user $n$ is active in coherence block $j-1$, then with probability $\alpha$, it is still active in coherence block $j$; if user $n$ is inactive in coherence block $j-1$, then with probability $\beta$, it is active in coherence block $j$. Given the above temporal correlation, over two consecutive coherence blocks $j-1$ and $j$, we have the following four cases to model each device's activity:

\emph{Case 1}: An user is active for both coherence blocks $j-1$ and $j$, i.e., $\boldsymbol{x}_n^{\left(j-1\right)}=\boldsymbol{h}_n^{\left(j-1\right)}$ and $\boldsymbol{x}_n^{\left(j\right)}=\boldsymbol{h}_n^{\left(j\right)}$, with probability $\alpha\lambda$.

\emph{Case 2}: An user is active at coherence block $j-1$, but becomes inactive at coherence block $j$, i.e., $\boldsymbol{x}_n^{\left(j-1\right)}=\boldsymbol{h}_n^{\left(j-1\right)}$ and $\boldsymbol{x}_n^{\left(j\right)}=\boldsymbol{0}$, with probability $(1-\alpha)\lambda$.

\emph{Case 3}: An user is inactive at coherence block $j-1$, but becomes active at coherence block $j$, i.e., $\boldsymbol{x}_n^{\left(j-1\right)}=\boldsymbol{0}$ and $\boldsymbol{x}_n^{\left(j\right)}=\boldsymbol{h}_n^{\left(j\right)}$, with probability $\beta(1-\lambda)$.

\emph{Case 4}: An user is inactive for both coherence blocks $j-1$ and $j$, i.e., $\boldsymbol{x}_n^{\left(j-1\right)}=\boldsymbol{0}$ and $\boldsymbol{x}_n^{\left(j\right)}=\boldsymbol{0}$, with probability $(1-\beta)(1-\lambda)$.

Similar to \cite{DCS-AMP,SI-AMP}, we assume that each Markov chain operates in steady-state such that the probability that a device becomes active is $\lambda$ over all the $J$ coherence blocks, i.e., (\ref{p_active}). Under this condition, the relation between $\alpha$ and $\beta$ is given by $\alpha\lambda+\beta(1-\lambda)=\lambda$. Due to this relation, the Markov chains are completely characterized by two parameters $\lambda$ and $\alpha$.

Under the temporal correlation modeled by (\ref{p_transition}), we should not detect the user activity over consecutive coherence blocks in an independent manner as in \cite{mMTC2, mMTC3}, since the user activity at the previous coherence block can provide SI for improving the estimation accuracy at the current coherence block. However, at each coherence block $j$, only an imperfect estimation of the device activity at coherence block $j-1$, denoted by $\hat{\delta}_n^{(j-1)}$, $\forall n$, is available at the BS. Despite the temporal correlation shown in (\ref{p_transition}), it is non-trivial to model a precise statistical relation between $\delta_n^{(j)}$ and $\hat{\delta}_n^{(j-1)}$, $\forall n$, since the connection between $\delta_n^{(j-1)}$ and $\hat{\delta}_n^{(j-1)}$, $\forall n$, is in general unknown. Without such a relation characterization, it is possible that the imperfect estimation at the previous coherence block is not properly utilized, which may even degrade the estimation performance at the current coherence block. This motivates us to study a systematic approach that is able to always leverage SI to improve the performance of activity detection and channel estimation.

Note that in the case without using SI, \cite{mMTC2} and \cite{mMTC3} showed that the estimation of $\boldsymbol{X}^{\left(j\right)}$ based on (\ref{basebandmodel}) is a compressed sensing problem, since many rows in $\boldsymbol{X}^{\left(j\right)}$ are zero vectors due to the sparse user activity. Moreover, the MMV-AMP algorithm has been used to estimate the row-sparse matrix $\boldsymbol{X}^{\left(j\right)}$ at each coherence block. In the rest of this paper, we study connection of SI to the current estimation and the method to embed SI into the AMP algorithm design.

\section{Leveraging SI in AMP}

This paper adopts the framework proposed in \cite{SI-AMP} to integrate SI into the MMV-AMP algorithm. At coherence block $j$, the SI-aided MMV-AMP algorithm will generate an estimation of $\boldsymbol{X}^{(j)}$, denoted by $\hat{\boldsymbol{X}}^{(j)}=[\hat{\boldsymbol{x}}_1^{(j)},\ldots,\hat{\boldsymbol{x}}_N^{(j)}]^T$, based on the signal received at the current time slot (\ref{basebandmodel}) and the estimation made by SI-aided MMV-AMP algorithm at the previous coherence block, i.e., $\hat{\boldsymbol{X}}^{(j-1)}$. Specifically, at coherence block $j$, the SI-aided MMV-AMP algorithm starts from $\boldsymbol{X}_0^{\left(j\right)}=\boldsymbol{0}$ and $\boldsymbol{R}_0^{\left(j\right)}=\boldsymbol{Y}^{\left(j\right)}$ and iterates as follows:
\begin{align}
&    \boldsymbol{x}_{n, t+1}^{\left(j\right)}= \eta_{n, t}^{\left(j\right)}\left ( \boldsymbol{x}_{n, t}^{\left(j\right)}+\left(\boldsymbol{R}_t^{\left(j\right)}\right)^H \boldsymbol{s}_{n},  f_{n,j}\left(\boldsymbol{\hat{x}}_n^{\left(j-1\right)}\right)\right), \label{amp_x} \\
&   \boldsymbol{R}_{t+1}^{\left(j\right)}\!=\!  \boldsymbol{Y}^{\left(j\right)}\!- \boldsymbol{S}\! \boldsymbol{X}_{t+1}^{\left(j\right)}\!  \nonumber \\ & ~~~~~~~ +\frac{N}{L}\! \boldsymbol{R}_{t}^{(j)}\! \left \langle\!  {\eta_{n, t}^{\left(j\right)}}'\! \left (\!\boldsymbol{x}_{n,t}^{\left(j\right)}\! +\! \left(\!\boldsymbol{R}_t^{\left(\!j\!\right)}\!\right)^H \!\!\!\boldsymbol{s}_{n}, \!f_{n,j}\! \left(\!\boldsymbol{\hat{x}}_n^{\left(\!j-1\!\right)}\!\right )\! \right)\! \right\rangle\!. \label{amp_r}
\end{align}
In (\ref{amp_x}) and (\ref{amp_r}), $t=0,1,\ldots$ denotes the index of algorithm iteration, $\boldsymbol{X}_{t}^{\left(j\right)} = [\boldsymbol{x}_{1, t}^{\left(j\right)}, \ldots, \boldsymbol{x}_{N, t}^{\left(j\right)}]^T$ denotes the estimation of $\boldsymbol{X}^{\left(j\right)}$ at the $t$-th iteration of the AMP algorithm, $f_{n,j}(\boldsymbol{\hat{x}}_n^{\left(j-1\right)})$ is a function of $\boldsymbol{\hat{x}}_n^{\left(j-1\right)}$ which is used as the SI for device $n$, $\boldsymbol{R}_t^{\left(j\right)}$ is the corresponding residual at iteration $t$, ${\eta}_{n, t}^{\left(j\right)} (\cdot, \diamond)\in\mathbb{C}^{M\times 1}$ is the denoising function for device $n$, ${\eta_{n, t}^{\left(j\right)}}' (\cdot, \diamond)$ is the first-order derivative of ${\eta}_{n, t}^{\left(j\right)}(\cdot, \diamond)$ with respect to the first variable $\cdot$, and $\left \langle \cdot  \right \rangle $ is the averaging operation over all entries of ${{\eta}_{n, t}^{\left(j\right)}}'( \cdot, \diamond )$. Let $\boldsymbol{X}_{\infty}^{\left(j\right)} = [\boldsymbol{x}_{1, \infty}^{\left(j\right)}, \ldots, \boldsymbol{x}_{N, \infty}^{\left(j\right)}]^T$ and $\boldsymbol{R}_{\infty}^{\left(j\right)}$ denote the estimation of $\boldsymbol{x}_{n}^{\left(j\right)}$ and the corresponding residual after the convergence of the SI-aided MMV-AMP algorithm at coherence block $j$. Then, we have $\hat{\boldsymbol{x}}_n^{(j-1)}=\boldsymbol{x}_{n,\infty}^{\left(j-1\right)}$, $\forall j,n$.

Different from the conventional MMV-AMP algorithm \cite{AMP,MMV_AMP} used in \cite{mMTC2,mMTC3}, the estimation at the previous coherence block is utilized for denoiser design as given by (\ref{amp_x}) under our considered SI-aided MMV-AMP algorithm. To implement this algorithm in IoT systems with temporally-correlated activity, in the following, we introduce what SI should be extracted from previous estimation, i.e., the design of $f_{n,j}(\boldsymbol{\hat{x}}_n^{\left(j-1\right)})$, and how to design the denoisers based on this SI.
\subsection{Identifying SI from State Evolution}
According to \cite{SI-AMP}, with the SI-aided MMV-AMP algorithm shown in (\ref{amp_x}) and (\ref{amp_r}), there exists the state evolution in the asymptotic regime where $N,L\to \infty$ with fixed $N/L$. Specifically, at each iteration $t$ of AMP to estimate $\boldsymbol{X}^{(j)}$, $ \boldsymbol{x}_{n,t}^{\left(j\right)}+\left(\boldsymbol{R}_t^{\left(j\right)}\right)^H \!\!\boldsymbol{s}_{n}$ is statistically equivalent to:
\vspace{-4pt}
\begin{equation}\label{model}
    \boldsymbol{\tilde{x}}_{n,t}^{\left(j\right)} = \boldsymbol{x}_n^{\left(j\right)}+\left ( \boldsymbol{\Sigma}_t^{\left(j\right)}\right )^\frac{1}{2}\boldsymbol{v}_n^{(j)}, ~~~ \forall n,j,
\vspace{-2pt}
\end{equation}
where $\boldsymbol{v}_n^{(j)}\in\mathbb{C}^{M\times 1}\sim\mathcal{CN}(\boldsymbol{0},\boldsymbol{I})$ is the noise independent of $\boldsymbol{x}_n^{\left(j\right)}$ and $\boldsymbol{\Sigma}_t^{\left(j\right)}\in \mathbb{C}^{M\times M}$ is the \emph{state}. Define a set of random vectors $\boldsymbol{X}_n^{\left(j\right)}\in\mathbb{C}^{M\times 1}$, $\boldsymbol{V}_n^{(j)}\in\mathbb{C}^{M\times 1}$, and $\boldsymbol{\hat{X}}_n^{\left(j-1\right)}\in\mathbb{C}^{M\times 1}$ which capture the distribution of $\boldsymbol{x}_n^{\left(j\right)}$, $\boldsymbol{v}_n^{(j)}$, and $\boldsymbol{\hat{x}}_n^{\left(j-1\right)}$, respectively, $\forall j,n$. The state evolution is given by:
\begin{equation}\label{state_evolution}
\begin{aligned}
    &\boldsymbol{\Sigma}_{t+1}^{\left(j\right)} = \sigma_z^2\boldsymbol{I}+\\
    &\frac{N}{L}\mathop{\mathbb{E}} \!\left [\! \left(\! \eta_{n,t}^{\left(j\right)}\left (\!  \boldsymbol{X}_n^{\left(j\right)}\!+\!\left (\!  \boldsymbol{\Sigma}_t^{\left(j\right)}\! \right  )^\frac{1}{2}\boldsymbol{V}_n^{(j)},  f_{n,j}\! \left(\! \boldsymbol{\hat{X}}_n^{\left(j-1\right)}\! \right)\! \right ) \!-\!\boldsymbol{X}_n^{\left(j\right)}\! \right)^{H}\right.\\
    &\qquad\left. \left(\!  \eta_{n,t}^{\left(j\right)}\!\left(\!  \boldsymbol{X}_n^{\left(j\right)}\!+\!\left (\boldsymbol{\Sigma}_t^{\left(j\right)}\right)^\frac{1}{2}\boldsymbol{V}_n^{(j)}, f_{n,j}\!\left(\! \boldsymbol{\hat{X}}_n^{\left(j-1\right)}\right)\!  \right) \! -\! \boldsymbol{X}_n^{\left(j\right)}\! \right)\! \right].
\end{aligned}
\end{equation}
Note that at coherence block $j$, we already have the estimation of $\boldsymbol{X}^{(j-1)}$, i.e., $\boldsymbol{\hat{x}}_n^{\left(j-1\right)}=\boldsymbol{x}_{n,\infty}^{\left(j-1\right)}$, $\forall n$. According to (\ref{model}), $\boldsymbol{\hat{x}}_{n}^{\left(\!j-1\!\right)}\!+\!\left(\!\boldsymbol{R}_\infty^{\left(\!j-1\!\right)}\!\right)^H \!\!\boldsymbol{s}_{n}$ is statistically equivalent to:
\begin{equation}\label{state_estimation}
   \boldsymbol{\tilde{x}}_{n,\infty}^{\left(j-1\right)}= \boldsymbol{x}_n^{\left(j-1\right)}+\left(\boldsymbol{\Sigma}_\infty^{\left(j-1\right)}\right)^{\frac{1}{2}}\boldsymbol{v}_n^{(j-1)}, ~~~ \forall n,j,
\end{equation}
where $\boldsymbol{\Sigma}_\infty^{\left(j-1\right)}$ denotes the state of the AMP algorithm (\ref{state_evolution}) after it converges at coherence block $j-1$. It is worth noting that (\ref{state_estimation}) reveals the correlation between $\boldsymbol{\tilde{x}}_{n,\infty}^{\left(j-1\right)}$ and $\boldsymbol{x}_n^{\left(j-1\right)}$, while (\ref{p_transition}) reveals the correlation between $\boldsymbol{x}_n^{\left(j-1\right)}$ and $\boldsymbol{x}_n^{\left(j\right)}$. Thus, the correlation between $\boldsymbol{\tilde{x}}_{n,\infty}^{\left(j-1\right)}$ and $\boldsymbol{x}_n^{\left(j\right)}$ can be built for all the devices. This motivates us to adopt the following SI to design the denoisers in the AMP algorithm:
\begin{equation}\label{SI}
f_{n,j}\left(\boldsymbol{\hat{x}}_n^{\left(j-1\right)}\right)=\boldsymbol{\hat{x}}_n^{\left(j-1\right)}+\left(\!\boldsymbol{R}_\infty^{\left(\!j-1\!\right)}\!\right)^H \!\!\boldsymbol{s}_{n}, ~~~ \forall n,j.
\end{equation}
The next question is how to utilize the correlation between $\boldsymbol{\hat{x}}_n^{\left(j-1\right)}\!+\!\left(\!\boldsymbol{R}_\infty^{\left(\!j-1\!\right)}\!\right)^H \!\!\boldsymbol{s}_{n}$'s and $\boldsymbol{x}_n^{\left(j\right)}$'s to design the denoisers (\ref{amp_x}).
\subsection{Leveraging SI for MMSE Denoiser Design}\label{subsec_sifordenoiser}

In this paper, we adopt the Baysian approach to design the MMSE denoisers $\eta_{n,t}^{\left(j\right)}\left ( \cdot, \diamond \right )$'s for signal recovery at each coherence block. At the $(t+1)$-th iteration of the AMP algorithm at coherence block $j$, the available information includes $ \boldsymbol{x}_{n,t}^{\left(j\right)}+\left(\boldsymbol{R}_t^{\left(j\right)}\right)^H \!\!\boldsymbol{s}_{n}$ from the current coherence block whose distribution is modeled by (\ref{model}) and the SI from the previous coherence block $ \boldsymbol{\hat{x}}_n^{\left(j-1\right)}+\left(\!\boldsymbol{R}_\infty^{\left(\!j-1\!\right)}\!\right)^H \!\!\boldsymbol{s}_{n}$ whose distribution is modeled by (\ref{state_estimation}). Based on the above information, the MMSE denoisers can be expressed as
\begin{align}\label{expformmse}
    \mathop{\mathbb{E}}[&\boldsymbol{X}_n^{\left(j\right)}\mid\boldsymbol{X}_n^{\left(j\right)}+\left ( \boldsymbol{\Sigma}_t^{\left(j\right)}\right )^\frac{1}{2}\boldsymbol{V}_n^{(j)} = \boldsymbol{\tilde{x}}_{n,t}^{\left(j\right)}, \nonumber \\
    &\boldsymbol{{X}}_n^{\left(j-1\right)}+\left ( \boldsymbol{\Sigma}_\infty^{\left(j-1\right)}\right )^\frac{1}{2}\boldsymbol{V}_n^{(j-1)}=\boldsymbol{\tilde{x}}_{n,\infty}^{\left(j-1\right)}], ~ \forall n,j.
\end{align}

Based on the similar approach used in \cite[Appendix B]{mMTC2}, it can be shown that with the above MMSE denoisers, the matrix $\boldsymbol{\Sigma}_t^{\left(j\right)}$ generated by the state evolution (\ref{state_evolution}) always stays as a scaled version of the identity matrix, i.e.,
\begin{equation}\label{dia}
    \boldsymbol{\Sigma}_t^{\left(j\right)} = \left(\tau_t^{\left(j\right)}\right)^2\boldsymbol{I}, ~~~ \forall t,j.
\end{equation}With this result, the MMSE denoisers (\ref{expformmse}) can be explicitly characterized by the following theorem.

\begin{figure}[t]
	\centering
	\includegraphics[height=5.5cm]{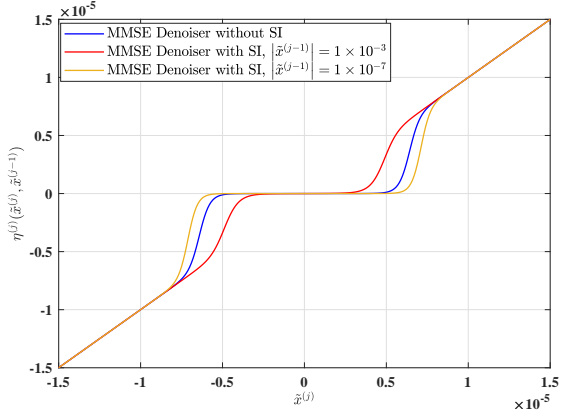}
	\vspace{-4pt}
	\caption{Comparison of MMSE denoisers with/without using SI.} \label{Fig_denoiser}
	\vspace{-10pt}
\end{figure}

\newtheorem{theorem}{Theorem}
\begin{theorem}\label{theorem1}
Consider the SI-aided MMV-AMP algorithm given by (\ref{amp_x}) and (\ref{amp_r}) under the temporal correlation model for user activity shown in (\ref{p_transition}). Define
\begin{align}
 \Delta_{n,t}^{\left(j\right)}=\left(\tau_{t}^{\left(j\right)}\right)^{-2}-\left(\left(\tau_t^{\left(j\right)}\right)^{2}+\gamma_n\right)^{-1}, ~ \forall n,t,j. \label{Delta1}
\end{align}
Under the asymptotic regime where $N,L\to \infty$ with fixed $N/L$, the MMSE denoisers (\ref{expformmse}) at coherence block $j$ with the SI given in (\ref{SI}) are expressed as:
\begin{equation}\label{simplified denoiser}
  \eta_{t,n}^{\!\left(\!j\!\right)\!} \!\left(\!\boldsymbol{\tilde{x}}_{n,t}^{\left(j\right)}, \boldsymbol{\tilde{x}}_{n,\infty}^{\left(j-1\right)}\!\right)\!
=\frac{\gamma_n\left(\gamma_n+\left(\tau_t^{\left(j\right)}\right)^2\right)^{-1}\boldsymbol{\tilde{x}}_{n,t}^{\left(j\right)}}{1+\frac{1-\lambda}{\lambda}\mu_{n,t}^{\left(j\right)}\times \frac{\beta+(1-\beta)\mu_{n,\infty}^{\left(j-1\right)} }{\alpha+(1-\alpha)\mu_{n,\infty}^{\left(j-1\right)}}}, ~ \forall n,t,j,
\end{equation}
where
\begin{equation}\label{mu}
\mu_{n,t}^{\left(j\right)}= \left(\frac{\left(\tau_t^{\left(j\right)}\right)^2+\gamma_n}{\left(\tau_t^{\left(j\right)}\right)^2}\right)^M{\rm exp}\left(-\Delta_{n,t}^{\left(j\right)}\left\|\boldsymbol{\tilde{x}}_{n,t}^{\left(j\right)}\right\|^2\right),
\end{equation}and $(\tau_{\infty}^{(j-1)})^2$ can be obtained from the state evolution (\ref{state_evolution}) and (\ref{dia}) after AMP converges in coherence block $j-1$.
\end{theorem}
\begin{IEEEproof}
	Please refer to Appendix \ref{appa}.
\end{IEEEproof}

To gain insights from Theorem \ref{theorem1}, we consider some special cases. First, if user activity is independent over different coherence blocks, i.e., $\alpha=\beta=\lambda$ such that $Pr(\delta_n^{\left(j\right)}\mid \delta_n^{\left(j-1\right)}) =Pr(\delta_n^{\left(j\right)})$, $\forall n,j$, the denoisers (\ref{simplified denoiser}) will reduce to
\begin{equation}\label{denoiserwithoutsi}
 \eta_{n,t}^{\left(j\right)} \!\left(\!\boldsymbol{\tilde{x}}_{n,t}^{\left(j\right)}, \boldsymbol{\tilde{x}}_{n,\infty}^{\left(j-1\right)}\!\right)\!
\!=\!\frac{\gamma_n\!\left(\!\gamma_n\!+\!(\tau_t^{(j)})^2\!\right)^{-1}\!\boldsymbol{\tilde{x}}_{n,t}^{\left(j\right)}}{1+\frac{1-\lambda}{\lambda}\mu_{n,t}^{\left(j\right)}}, ~ \forall n,j,
\end{equation}
which are the MMSE denoisers proposed in \cite{mMTC2}\cite{mMTC3} without taking SI into account. This is because if there is no temporal correlation in user activity, SI will have no effect on the MMSE denoiser design. Second, if $(\tau_{\infty}^{(j-1)})^2\to\infty$, it can be shown from (\ref{mu}) that $\mu_{n,\infty}^{\left(j-1\right)}=1$, $\forall n$. Then, the MMSE denoisers shown in (\ref{simplified denoiser}) will also reduce to the MMSE denoisers (\ref{denoiserwithoutsi}) proposed in \cite{mMTC2}\cite{mMTC3} without taking SI into account. This is because according to (\ref{state_estimation}) and (\ref{dia}), $(\tau_{\infty}^{(j-1)})^2$ can be viewed as the equivalent noise power for estimating $\boldsymbol{X}^{(j-1)}$ by AMP. If this noise power is infinite but the power of each row in $\boldsymbol{X}^{(j-1)}$ is finite, then the estimation does not provide any useful information for the estimation in the next block, despite the existence of temporal correlation in activity.

Next, we provide a numerical example to compare the denoisers using SI, i.e., (\ref{simplified denoiser}), and without using SI, i,e., (\ref{denoiserwithoutsi}). In this example, we set $M=1$, $\lambda=0.1$, $\alpha=0.91$, $\beta=0.01$, $\tau^{\left(j\right)}=\tau^{\left(j-1\right)}=2\times10^{-6}$, and $\gamma=\gamma=1\times10^{-8}$. Fig. 1 shows the SI-aided MMSE denoisers when $|\tilde{x}^{\left(j-1\right)}|=1\times10^{-3}$ and $|\tilde{x}^{\left(j-1\right)}|=1\times10^{-7}$ as well as the denoiser without using SI. Compared to the denoiser without using SI, it is observed that when $|\tilde{x}^{\left(j-1\right)}|$ is larger/smaller, i.e., the user tends to be detected as an active/inactive device at the previous block, the SI-aided MMSE denoiser estimates $x^{(j)}$ as zero over a smaller/larger range of $\tilde{x}^{(j)}$, i.e., the user tends to be detected as an active/inactive device at the current time slot.

\begin{figure}[t]
	\centering
	\includegraphics[height=5.5cm]{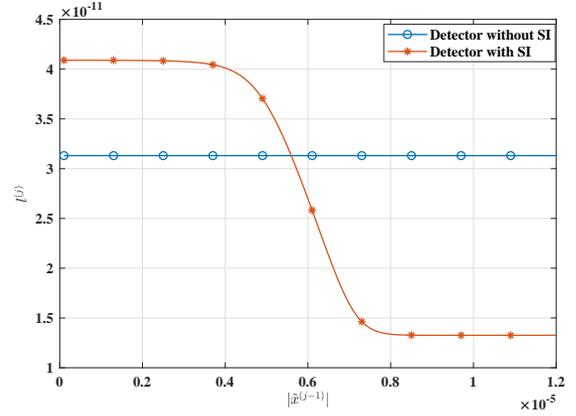}
    \vspace{-4pt}
	\caption{Comparison of threshold for activity detectors with/without using SI.} \label{Fig_detector}
    \vspace{-10pt}
\end{figure}

\subsection{Leveraging SI for Activity Detector Design}

After the convergence of the SI-aided MMV-AMP \mbox{algorithm}, the LLR test is applied to conduct device activity detection. For the hypothesis detection problem $H_0$: user is inactive; $H_1$: user is active, the LLR-based detector is
\begin{equation}\label{LLR}
        LLR_{n,t}^{(j)}=\log\left (\frac{p(\boldsymbol{\tilde{x}}_{n,t}^{(j)},\boldsymbol{\tilde{x}}_{n,\infty}^{(j-1)}\mid \boldsymbol{{x}}_{n}^{(j)}\ne \boldsymbol{0})}{p(\boldsymbol{\tilde{x}}_{n,t}^{(j)},\boldsymbol{\tilde{x}}_{n,\infty}^{(j-1)}\mid \boldsymbol{{x}}_{n}^{(j)} =\boldsymbol{0})} \right ) \overset{H_1}{\underset{H_0}{\gtrless}} l,
\end{equation}
where $l$ is a common threshold of LLR for all the users over all the coherence blocks.

\begin{theorem}\label{theorem2}
Consider the SI-aided MMV-AMP algorithm given by (\ref{amp_x}) and (\ref{amp_r}) under the temporal correlation model for user activity shown in (\ref{p_transition}). Under the asymptotic regime where $N,L\to \infty$ with fixed $N/L$, the LLR-based decision rule (\ref{LLR}) can be expressed as:
\begin{equation}\label{simplifieddetector}
\begin{aligned}
\|\boldsymbol{\tilde{x}}_{n,t}^{(j)}\|^2\overset{H_1}{\underset{H_0}{\gtrless}}l_{n,t}^{\left(j\right)}&\triangleq\frac{1}{\Delta_{n,t}^{\left(j\right)}}\!\bigg(\!l+M\log\!\bigg(\!\frac{\!(\!\tau_t^{(j)}\!)^2\!+\!\gamma_n\!}{\!(\!\tau_t^{(j)}\!)^2}\!\bigg)\!+\\&\log\bigg(\frac{\beta+(1-\beta)\mu_{n,\infty}^{(j-1)}}{\alpha+(1-\alpha)\mu_{n,\infty}^{(j-1)}}\bigg)\!\bigg)\!, ~ \forall n,t,j,
\end{aligned}
\end{equation}
where $\Delta_{n,t}^{\left(j\right)}$ and $\mu_{n,\infty}^{(j-1)}$ are given by (\ref{Delta1}) and (\ref{mu}), respectively.
\end{theorem}
\begin{IEEEproof}
Please refer to Appendix B.
\end{IEEEproof}

Theorem \ref{theorem2} states that a device is detected to be active if $\|\boldsymbol{\tilde{x}}_{n,t}^{(j)}\|^2$ is larger than a threshold, which depends on the SI from the previous block. Similar to the MMSE denoisers in Theorem \ref{theorem1}, if $\alpha=\beta=\lambda$, or if $\tau_\infty^{(j-1)}\to \infty$, then the LLR-based detectors (\ref{simplifieddetector}) will reduce to:
\begin{equation}\label{l_withoutsi}
    \|\boldsymbol{\tilde{x}}_{n,t}^{(j)}\|^2\overset{H_1}{\underset{H_0}{\gtrless}} \frac{l+M\log\left(\frac{\left(\tau_t^{(j)}\right)^2+\gamma_n}{\left(\tau_t^{(j)}\right)^2}\right)}{\Delta_{n,t}^{\left(j\right)}}, ~ \forall n,t,j,
\end{equation}
which are the detectors without using SI \cite{mMTC3}.

Next, we provide a numerical example to compare the LLR-based activity detectors using SI, i.e., (\ref{simplifieddetector}), and those without using SI, i.e., (\ref{l_withoutsi}). The setup is the same as that for Fig. \ref{Fig_denoiser}. Moreover, we set $l=0$. Fig. \ref{Fig_detector} shows the threshold in the activity detectors, i.e., $l^{\left(j\right)}$, versus different values of $|\tilde{x}^{(j-1)}|$, which is the SI from the previous time slot. Compared to the case without using SI, it is observed that if $|\tilde{x}^{(j-1)}|$ is larger/smaller, i.e., the user tends to be detected as an active/inactive user previously, the threshold in the SI-aided detectors (\ref{simplifieddetector}) becomes smaller/larger, i.e., this user tends to be detected as an active/inactive user at the current time slot.

\vspace{-2pt}
\section{Numerical Results}
\vspace{-2pt}
In this section, we provide numerical results to evaluate the performance of the proposed SI-aided MMV-AMP algorithm in massive IoT connectivity systems. We assume that there are $N=4000$ devices randomly located in a cell of radius $R=1000$m. The path loss model is $-128.1-36.7\log_{10}(d_n)$ in dB, where $d_n$ in km denotes the distance from device $n$ to the BS. We consider the communication over $J=10$ coherence blocks, while at each time slot, we have  $\lambda=0.1$, $\alpha=0.46$, and $\beta=0.06$. Next, the user transmit power is 23 dBm. Last, the power spectrum density of the noise is -169 dBm/Hz, while the bandwidth of the channel is assumed to be 10 MHz.

\begin{figure}[t]
	\centering
	\includegraphics[height=6.6cm]{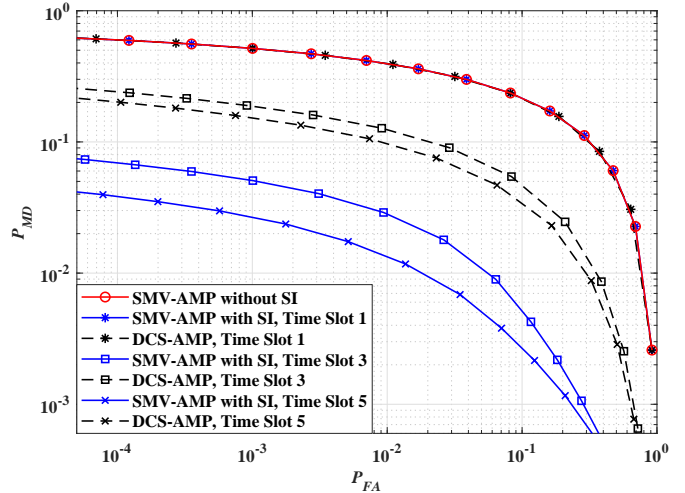}
	\vspace{-4pt}
	\caption{Activity detection under SMV-AMP algorithm with SI.} \label{compare}
	\vspace{-4pt}
\end{figure}

First, we consider the case when the BS is equipped with one antenna, i.e., $M=1$. In this case, we call our proposed algorithm as SMV-AMP with SI. In Fig. \ref{compare}, we show the tradeoff between the probabilities of false alarm ${P_{FA}}$ and missed detection ${P_{MD}}$, which is obtained by varying the value of $l$ in the activity detectors. In this numerical example, we set $L=600$. Moreover, we consider the Dynamic Compressed Sensing via Approximate Message Passing (DCS-AMP) algorithm proposed in \cite{DCS-AMP} as the benchmark scheme. The DCS-AMP is implemented in filtering mode to match our setting. It is observed from Fig. \ref{compare} that under our proposed SMV-AMP algorithm with SI, the activity detection performance improves over time. This shows that the
proposed algorithm is capable of intelligently exploiting the SI obtained in the previous time slots for improving the detection performance. Moreover, at time slot 5, our proposed scheme can achieve much lower detection error probability than the conventional AMP algorithm without utilizing SI. It is also observed that the proposed SI-aided SMV-AMP algorithm outperforms the DCS-AMP algorithm significantly at time slots 3 and 5 because our proposed scheme is built on the true statistical correlation between the  estimated activity in the last time slot and true activity in the current time slot.

\begin{figure}[t]
	\centering
	\includegraphics[height=6.6cm]{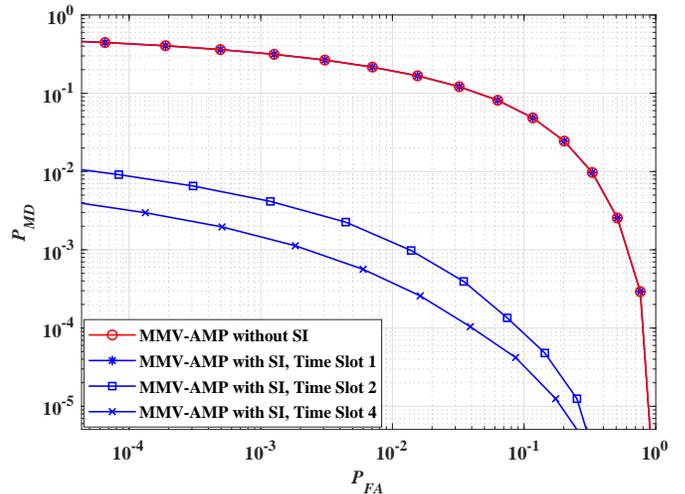}
	\vspace{-4pt}
	\caption{Activity detection under MMV-AMP algorithm with SI.} \label{m2}
	\vspace{-4pt}
\end{figure}

Next, we consider the case when the BS is equipped with multiple antennas. In this case, we term our proposed algorithm as MMV-AMP with SI. Fig. 4 shows the tradeoff between the false alarm probability $P_{FA}$ and missed detection probability $P_{MD}$ when $M=2$ and $L=500$. It is observed that the proposed algorithm with SI achieves significant performance gain over the one without using SI; moreover, the performance improves as the time slot index increases. These results in Fig. \ref{compare} and Fig. \ref{m2} show that by smartly leveraging SI, the proposed algorithm is able to achieve satisfactory detection performance with only a small number of BS antennas (e.g., even 1 or 2), thus being a promising cost-effective solution for future massive IoT systems.


\section{Conclusion}
In this paper, we utilized the temporal correlation in user activity for device activity detection in IoT systems. The main motivation is to achieve satisfactory detection performance with a smaller number of BS antennas and thus lower computational complexity. Along this line, we designed a framework of SI-aided MMV-AMP, where the estimation at the previous time slot was leveraged as SI to devise better MMSE denoisers and activity detectors at the current time slot.

\begin{appendices}
\section{Proof of Theorem\ref{theorem1}}\label{appa}
In this proof, for simplicity, we omit the subscripts $t$ and $n$ in the all the notations. Then, in the MMSE denoisers (\ref{expformmse}), the conditional expectation can be given by
\begin{equation}\label{expectation}
\begin{aligned}
&E[\boldsymbol{X}^{\left(j\right)}\mid \boldsymbol{\tilde x}^{\left(j\right)}, \boldsymbol{\tilde x}_\infty^{\left(j-1\right)}]\\
\overset{{\rm (a)}}{=}&E[\boldsymbol{X}^{\left(j\right)} \mid \boldsymbol{\tilde x}^{\left(j\right)}, \boldsymbol{\tilde{x}}_\infty^{(j-1)},Case ~ 1]p\left( {Case ~ 1}\mid \boldsymbol{\tilde x}^{\left(j\right)}, \boldsymbol{\tilde x}_\infty^{\left(j-1\right)}\right)\\
&+E[\boldsymbol{X}^{\left(j\right)} \mid \boldsymbol{\tilde x}^{\left(j\right)}, \boldsymbol{\tilde x}_\infty^{\left(j-1\right)},Case ~ 3]p\left( {Case ~ 3}\mid \boldsymbol{\tilde x}^{\left(j\right)}, \boldsymbol{\tilde{x}}_\infty^{\left(j-1\right)}\right),
\end{aligned}
\end{equation}
where (a) is because $\boldsymbol{X}^{\left(j\right)}=\boldsymbol{0}$ for \emph{Case 2} and \emph{Case 4} according to Section \ref{sec_sys}. In the following, we focus on \emph{Case 1} and \emph{Case 3} to characterize (\ref{expectation}). 

\emph{Case 1}:
According to Section \ref{sec_sys}, under \emph{Case 1}, it follows that $\boldsymbol{x}_n^{\left(j-1\right)}=\boldsymbol{h}_n^{\left(j-1\right)}$ and $\boldsymbol{x}_n^{\left(j\right)}=\boldsymbol{h}_n^{\left(j\right)}$. Based on (\ref{model}), (\ref{state_estimation}), and (\ref{dia}), we have $\boldsymbol{\tilde x}^{\left(j\right)}=\boldsymbol{h}^{\left(j\right)}+\tau^{\left(j\right)}\boldsymbol{v}^{(j)}, \boldsymbol{\tilde x}_\infty^{\left(j-1\right)}=\boldsymbol{h}^{\left(j-1\right)}+\tau_\infty^{\left(j-1\right)}\boldsymbol{v}^{(j-1)}$. Thus, $E[\boldsymbol{X}^{\left(j\right)} \mid \boldsymbol{\tilde x}^{\left(j\right)}, \boldsymbol{\tilde{x}}^{\left(j-1\right)},Case ~ 1]$ in (\ref{expectation}) can be given by
\begin{equation}\label{exp_case1}
\begin{aligned}
    E[\boldsymbol{h}^{\left(j\right)} \mid & \boldsymbol{\tilde x}^{\left(j\right)}=\boldsymbol{h}^{\left(j\right)}+\tau^{\left(j\right)}\boldsymbol{v}^{(j)},\\ &\boldsymbol{\tilde{x}}_\infty^{\left(j-1\right)}=\boldsymbol{h}^{\left(j-1\right)}+\tau _\infty^{\left(j-1\right)}\boldsymbol{v}^{(j-1)}].
\end{aligned}
\end{equation}
Because $\boldsymbol{h}^{\left(j\right)}$, $\boldsymbol{h}^{\left(j-1\right)}$, $\boldsymbol{v}^{(j)}$, and $\boldsymbol{v}^{(j-1)}$ are independent with each other, we have
\begin{align}
& E[\boldsymbol{h}^{\left(j\right)} \mid  \boldsymbol{\tilde x}^{\left(j\right)}=\boldsymbol{h}^{\left(j\right)}+\tau^{\left(j\right)}\boldsymbol{v}^{(j)}, \nonumber \\ &~~~~~ \boldsymbol{\tilde{x}}_\infty^{\left(j-1\right)}=\boldsymbol{h}^{\left(j-1\right)}+\tau _\infty^{\left(j-1\right)}\boldsymbol{v}^{(j-1)}] \nonumber \\
= & E[\boldsymbol{h}^{\left(j\right)} \mid \boldsymbol{\tilde x}^{\left(j\right)}=\boldsymbol{h}^{\left(j\right)}+\tau^{\left(j\right)}\boldsymbol{v}^{(j)}] \\ =&{\gamma}\left(\gamma+\left(\tau^{\left(j\right)}\right)^2\right)^{-1}{\boldsymbol{\tilde x}^{\left(j\right)}}. \label{ecase1}
\end{align}

Next, we calculate $p\left({Case ~ 1}\mid \boldsymbol{\tilde x}^{\left(j\right)}, \boldsymbol{\tilde{x}}_\infty^{\left(j-1\right)}\right)$ in (\ref{expectation}) as follows
\begin{align}
&p\left( {Case ~ 1}\mid \boldsymbol{\tilde x}^{\left(j\right)},\boldsymbol{\tilde x}_\infty^{\left(j-1\right)}\right)\nonumber \\ 
=&\frac{p(\boldsymbol{\tilde x}^{\left(j\right)}, \boldsymbol{\tilde{x}}_\infty^{\left(j-1\right)}, Case ~ 1)}{p(\boldsymbol{\tilde x}^{\left(j\right)}, \boldsymbol{\tilde{x}}_\infty^{\left(j-1\right)})} \\ =&\frac{P(Case ~ 1)p(\boldsymbol{\tilde x}^{\left(j\right)}, \boldsymbol{\tilde{x}}_\infty^{\left(j-1\right)}\mid Case ~ 1)}{p(\boldsymbol{\tilde x}^{\left(j\right)}, \boldsymbol{\tilde{x}}_\infty^{\left(j-1\right)})} \\
=&\frac{\alpha\lambda\boldsymbol{\psi}_{\gamma+\left(\tau^{\left(j\right)}\right)^2}(\boldsymbol{\tilde x}^{\left(j\right)})\boldsymbol{\psi}_{\gamma+\left(\tau_\infty^{\left(j-1\right)}\right)^2}(\boldsymbol{\tilde{x}}_\infty^{\left(j-1\right)})}{p(\boldsymbol{\tilde x}^{\left(j\right)}, \boldsymbol{\tilde{x}}_\infty^{\left(j-1\right)})}, \label{pcase1}
\end{align}
where 
\begin{align}
\boldsymbol{\psi}_{\sigma^2}(\boldsymbol{x})=\frac{1}{\pi\left |(\sigma^2\boldsymbol{I}) \right | }{\rm exp}\left(-\frac{\boldsymbol{x}^H\boldsymbol{x}}{\sigma^2}\right),
\end{align}is the power density function (pdf) of a complex Gaussian random vector with zero mean and covariance $\sigma^2 \boldsymbol{I}$. We will derive the joint pdf $p(\boldsymbol{\tilde x}^{\left(j\right)},\boldsymbol{\tilde{x}}_\infty^{\left(j-1\right)})$ in (\ref{pcase1}) later.

\emph{Case 3}:
Similar to (\ref{ecase1}) in \emph{Case 1}, under \emph{Case 3}, it can be shown that
\begin{align}
& E[\boldsymbol{X}^{\left(j\right)} \mid \boldsymbol{\tilde x}^{\left(j\right)}, \boldsymbol{\tilde{x}}_\infty^{\left(j-1\right)},Case ~ 3] \nonumber \\ =&{\gamma}\left({\gamma+\left(\tau^{\left(j\right)}\right)^2}\right)^{-1}{\boldsymbol{\tilde x}^{\left(j\right)}}. \label{ecase3}
\end{align}
Moreover, similar to (\ref{pcase1}), it follows that
\begin{align}
&p\left( {Case ~ 3}\mid \boldsymbol{\tilde x}^{\left(j\right)}, \boldsymbol{\tilde{x}}_\infty^{\left(j-1\right)}\right)\nonumber \\ =&\frac{p(\boldsymbol{\tilde x}^{\left(j\right)}, \boldsymbol{\tilde{x}}_\infty^{\left(j-1\right)}, Case ~ 3)}{p(\boldsymbol{\tilde x}^{\left(j\right)}, \boldsymbol{\tilde{x}}_\infty^{\left(j-1\right)})} \\=&\frac{\beta(1-\lambda)\boldsymbol{\psi}_{\gamma+\left(\tau^{\left(j\right)}\right)^2}(\boldsymbol{\tilde x}^{\left(j\right)})\boldsymbol{\psi}_{\left(\tau_\infty^{\left(j-1\right)}\right)^2}(\boldsymbol{\tilde{x}}_\infty^{\left(j-1\right)}) }{p(\boldsymbol{\tilde x}^{\left(j\right)}, \boldsymbol{\tilde{x}}_\infty^{\left(j-1\right)})}. \label{pcase3}
\end{align}

To derive $p\left( {Case ~ 1}\mid \boldsymbol{\tilde x}^{\left(j\right)},\boldsymbol{\tilde x}_\infty^{\left(j-1\right)}\right)$ in (\ref{pcase1}) and $p\left( {Case ~ 3}\mid \boldsymbol{\tilde x}^{\left(j\right)}, \boldsymbol{\tilde{x}}^{\left(j-1\right)}\right)$ in (\ref{pcase3}), the last step is to characterize $p(\boldsymbol{\tilde x}^{\left(j\right)}, \boldsymbol{\tilde{x}}_\infty^{\left(j-1\right)})$. Similar to $p(\boldsymbol{\tilde x}^{\left(j\right)}, \boldsymbol{\tilde{x}}_\infty^{\left(j-1\right)}, Case ~ 1)$ shown in (\ref{pcase1}) and $p(\boldsymbol{\tilde x}^{\left(j\right)}, \boldsymbol{\tilde{x}}_\infty^{\left(j-1\right)}, Case ~ 3)$ shown in (\ref{pcase3}), it can be shown that
\begin{align}
    &p(\boldsymbol{\tilde x}^{\left(j\right)}, \boldsymbol{\tilde{x}}_\infty^{\left(j-1\right)},Case ~ 2)\notag\\=&(1-\alpha)\lambda \boldsymbol{\psi}_{\left(\tau^{\left(j\right)}\right)^2}(\boldsymbol{\tilde x}^{\left(j\right)})\boldsymbol{\psi}_{\gamma+\left(\tau_\infty^{\left(j-1\right)}\right)^2}(\boldsymbol{\tilde{x}}_\infty^{\left(j-1\right)}),\\
    &p(\boldsymbol{\tilde x}^{\left(j\right)}, \boldsymbol{\tilde{x}}_\infty^{\left(j-1\right)},Case ~ 4)\notag\\=&(1-\beta)(1-\lambda) \boldsymbol{\psi}_{\left(\tau^{\left(j\right)}\right)^2}(\boldsymbol{\tilde x}^{\left(j\right)})\boldsymbol{\psi}_{\left(\tau_\infty^{\left(j-1\right)}\right)^2}(\boldsymbol{\tilde{x}}_\infty^{\left(j-1\right)}).
\end{align}
Thus, it follows that
\begin{align}
&p(\boldsymbol{\tilde x}^{\left(j\right)}, \boldsymbol{\tilde{x}}^{\left(j-1\right)})\nonumber \\=& p(\boldsymbol{\tilde x}^{\left(j\right)}, \boldsymbol{\tilde{x}}^{\left(j-1\right)}, Case ~ 1)+p(\boldsymbol{\tilde x}^{\left(j\right)}, \boldsymbol{\tilde{x}}^{\left(j-1\right)}, Case ~ 2)\nonumber \\&+p(\boldsymbol{\tilde x}^{\left(j\right)}, \boldsymbol{\tilde{x}}^{\left(j-1\right)}, Case ~ 3)+p(\boldsymbol{\tilde x}^{\left(j\right)}, \boldsymbol{\tilde{x}}^{\left(j-1\right)}, Case ~ 4)\\
=&\alpha\lambda\boldsymbol{\psi}_{\gamma+\left(\tau^{\left(j\right)}\right)^2}(\boldsymbol{\tilde x}^{\left(j\right)})\boldsymbol{\psi}_{\gamma+\left(\tau_\infty^{\left(j-1\right)}\right)^2}(\boldsymbol{\tilde{x}}_\infty^{\left(j-1\right)})\nonumber \\&+(1-\alpha)\lambda \boldsymbol{\psi}_{\left(\tau^{\left(j\right)}\right)^2}(\boldsymbol{\tilde x}^{\left(j\right)})\boldsymbol{\psi}_{\gamma+\left(\tau_\infty^{\left(j-1\right)}\right)^2}(\boldsymbol{\tilde{x}}_\infty^{\left(j-1\right)}) \nonumber \\
&+\beta(1-\lambda)\boldsymbol{\psi}_{\gamma+\left(\tau^{\left(j\right)}\right)^2}(\boldsymbol{\tilde x}^{\left(j\right)})\boldsymbol{\psi}_{\left(\tau_\infty^{\left(j-1\right)}\right)^2}(\boldsymbol{\tilde{x}}_\infty^{\left(j-1\right)}) \nonumber \\
&+(1-\beta)(1-\lambda) \boldsymbol{\psi}_{\left(\tau^{\left(j\right)}\right)^2}(\boldsymbol{\tilde x}^{\left(j\right)})\boldsymbol{\psi}_{\left(\tau_\infty^{\left(j-1\right)}\right)^2}(\boldsymbol{\tilde{x}}_\infty^{\left(j-1\right)}). \label{totalp}
\end{align}
By plugging (\ref{ecase1}), (\ref{pcase1}), (\ref{ecase3}), (\ref{pcase3}), and (\ref{totalp}) into (\ref{expectation}), it can be shown that the MMSE denoisers by taking SI into account are expressed by (\ref{simplified denoiser}). Theorem \ref{theorem1} is thus proved.

\section{Proof of Theorem\ref{theorem2}}\label{appb}
It can be shown that
\begin{align}
   & p(\boldsymbol{\tilde{x}}_{n,t}^{\left(j\right)},\boldsymbol{\tilde{x}}_{n,\infty}^{\left(j-1\right)}\mid \boldsymbol{{x}}_{n}^{\left(j\right)}\ne \boldsymbol{0}) \nonumber \\ = & \frac{p(\boldsymbol{\tilde{x}}_{n,t}^{\left(j\right)},\boldsymbol{\tilde{x}}_{n,\infty}^{\left(j-1\right)}, {Case ~ 1})+p(\boldsymbol{\tilde{x}}_{n,t}^{\left(j\right)},\boldsymbol{\tilde{x}}_{n,\infty}^{\left(j-1\right)}, {Case ~ 3})}{p(\boldsymbol{{x}}_{n}^{\left(j\right)}\ne \boldsymbol{0})} \\ =& \frac{p(\boldsymbol{\tilde{x}}_{n,t}^{\left(j\right)},\boldsymbol{\tilde{x}}_{n,\infty}^{\left(j-1\right)}\mid {Case ~ 1})p({Case ~ 1})}{{p(\boldsymbol{{x}}_{n}^{\left(j\right)}\ne \boldsymbol{0})}} \nonumber \\ &+\frac{p(\boldsymbol{\tilde{x}}_{n,t}^{\left(j\right)},\boldsymbol{\tilde{x}}_{n,\infty}^{\left(j-1\right)}\mid {Case ~ 3})p({Case ~ 3})}{p(\boldsymbol{{x}}_{n}^{\left(j\right)}\ne \boldsymbol{0})} \\ = & 
    \boldsymbol{\psi}_{\gamma_n+(\tau_t^{\left(j\right)})^2}\! \left (\boldsymbol{\tilde{x}}_{n,t}^{\left(j\right)}\right )\times \nonumber \\ &\left(\! \alpha\boldsymbol{\psi}_{\gamma_n+(\tau_\infty^{\left(j-1\right)})^2}\!\left (\!  \boldsymbol{\tilde{x}}_{n,\infty}^{\left(j-1\right)} \! \right ) \!+\! (1-\alpha)\boldsymbol{\psi}_{(\tau_\infty^{\left(j-1\right)})^2}\! \left (\!  \boldsymbol{\tilde{x}}_{n,\infty}^{\left(j-1\right)} \! \right ) \right), \label{c1}
\end{align}where (\ref{c1}) is due to (\ref{pcase1}) and (\ref{pcase3}). Similarly, it can be shown that
\begin{align}
   & p(\boldsymbol{\tilde{x}}_{n,t}^{\left(j\right)},\boldsymbol{\tilde{x}}_{n,\infty}^{\left(j-1\right)}\mid \boldsymbol{{x}}_{n}^{\left(j\right)}= \boldsymbol{0}) =
    \boldsymbol{\psi}_{(\tau_t^{\left(j\right)})^2}\left (\boldsymbol{\tilde{x}}_{n,t}^{\left(j\right)}\right )\times \nonumber \\ &\left(\!\beta\boldsymbol{\psi}_{\gamma_n+(\tau_\infty^{\left(j-1\right)})^2}\! \left (\!  \boldsymbol{\tilde{x}}_{n,\infty}^{\left(j-1\right)} \! \right ) \! +\! (1-\beta)\boldsymbol{\psi}_{(\tau_\infty^{\left(j-1\right)})^2}\! \left (\!  \boldsymbol{\tilde{x}}_{n,\infty}^{\left(j-1\right)}\! \right )\right). \label{c2}
\end{align}
Therefore, by taking (\ref{c1}) and (\ref{c2}) into (\ref{LLR}), it follows that
\begin{align}
LLR_{n,t}^{(j)} & =\log\left(\frac{\boldsymbol{\psi}_{\gamma_n+(\tau_t^{\left(j\right))^2}}\left (\boldsymbol{\tilde{x}}_{n,t}^{\left(j\right)}\right )}{\boldsymbol{\psi}_{(\tau_t^{\left(j\right)})^2}\left (\boldsymbol{\tilde{x}}_{n,t}^{\left(j\right)}\right )}\right) \nonumber \\
&+\log\left(\frac{\alpha+(1-\alpha)\frac{\boldsymbol{\psi}_{(\tau_\infty^{\left(j-1\right)})^2}\left ( \boldsymbol{\tilde{x}}_{n,\infty}^{\left(j-1\right)}\right )}{\boldsymbol{\psi}_{\gamma_n+(\tau_\infty^{\left(j-1\right)})^2}\left ( \boldsymbol{\tilde{x}}_{n,\infty}^{\left(j-1\right)} \right )}}{\beta+(1-\beta)\frac{\boldsymbol{\psi}_{(\tau_\infty^{\left(j-1\right)})^2}\left ( \boldsymbol{\tilde{x}}_{n,\infty}^{\left(j-1\right)}\right )}{\boldsymbol{\psi}_{\gamma_n+(\tau_\infty^{\left(j-1\right)})^2}\left ( \boldsymbol{\tilde{x}}_{n,\infty}^{\left(j-1\right)} \right )}}\right).\label{llr2}
\end{align}With (\ref{llr2}), it can be shown that the detection rule (\ref{LLR}) is equivalent to (\ref{simplifieddetector}). Theorem \ref{theorem2} is thus proved.

\end{appendices}

\end{document}